%% file: main.tex
\documentclass[conference, onecolumn]{IEEEtran}
\IEEEoverridecommandlockouts
\usepackage{algorithmic}
\usepackage{amsmath,amssymb,amsfonts}
\usepackage{cite}
\usepackage{graphicx}
\usepackage{hyperref} 
\usepackage{listings}
\usepackage{multirow}
\usepackage{textcomp}
\usepackage{todonotes}
\usepackage{verbatim}
\usepackage{xcolor}
\usepackage{xspace} 

\def\c3sr{{C\textsuperscript{3}SR}\xspace}
\def\myscope{SCOPE}
\def\examplescope{Example$\vert$Scope\xspace}
\def\commscope{Comm$\vert$Scope\xspace}

\title{\myscope: \c3sr Systems Characterization and Benchmarking Framework\\
\thanks{
This work is supported by IBM-ILLINOIS Center for Cognitive Computing Systems Research (C3SR) - a research collaboration as part of the IBM AI Horizons Network.
}
}

\author{
\IEEEauthorblockN{Carl Pearson, Abdul Dakkak, Cheng Li, Sarah Hashash}
\IEEEauthorblockA{
    \textit{University of Illinois Urbana-Champaign}\\
    Urbana, IL, United States of America \\
    \texttt{\{\href{mailto:pearson@illinois.edu}{pearson},
    \href{mailto:dakkak@illinois.edu}{dakkak}, 
    \href{mailto:cli99@illinois.edu}{cli99},
    \href{mailto:hashash2@illinois.edu}{hashash2}\}@illinois.edu}
}
\and
\IEEEauthorblockN{Jinjun Xiong}
\IEEEauthorblockA{
    \textit{IBM T. J. Watson Research Center}\\
    Yorktown Heights, NY, United States of America \\
    \texttt{\href{mailto:jinjun@us.ibm.com}{jinjun@us.ibm.com}}
}
\and
\IEEEauthorblockN{Wen-mei Hwu}
\IEEEauthorblockA{
    \textit{University of Illinois Urbana-Champaign}\\
    Urbana, IL, United States of America \\
    \texttt{\href{mailto:w-hwu@illinois.edu}{w-hwu@illinois.edu}}
}
}

\begin{document}
\maketitle

\begin{abstract}
\input{abs}
\end{abstract}

\begin{IEEEkeywords}
benchmark, system, CUDA, python
\end{IEEEkeywords}

\input{sections/1-vision}
\input{sections/2-design-philosophy}

\input{sections/3-scope}
\input{sections/4-scope-submodules}
\input{sections/5-scope-utilities}
\input{sections/6-dev-maint}
\input{sections/7-conclusion}


\section*{Acknowledgments}

The authors would like to acknowledge contribution and insight from the following people:
I-Hsin Chung (IBM T. J. Watson Research Center),
Sarah Hashash and Andrew Schuh (University of Illinois at Urbana-Champaign)

\bibliographystyle{IEEEtran}
\bibliography{IEEEabrv,main}


\end{document}

%% file: abs.tex
This report presents the design of the Scope infrastructure for extensible and portable benchmarking.
Improvements in high-performance computing systems rely on coordination across different levels of system abstraction.
Developing and defining accurate performance measurements is necessary at all levels of the system hierarchy, and should be as accessible as possible to developers with different backgrounds.
The Scope project aims to lower the barrier to entry for developing performance benchmarks by providing a software architecture that allows benchmarks to be developed independently, by providing useful C/C++ abstractions and utilities, and by providing a Python package for generating publication-quality plots of resulting measurements.

%% file: sections/1-vision.tex
\section{Motivation}
\label{sec:vision}

The IBM-ILLINOIS Center for Cognitive Computing Systems Research (\c3sr) is a long-term collaboration between IBM 
Research and
the University of Illinois at Urbana-Champaign
focused on developing advanced Cognitive Computing
and Artificial Intelligence (AI) systems
that are optimized across the vertical stacks of
AI solutions, software, and hardware systems.
In particular, C3SR aims to develop technologies to 
improve cognitive application developers' productivity on heterogeneous infrastructure \cite{c3sr2018}.

This effort demands us to perform various system 
characterization and performance measurements work
at all levels of computer system abstraction, 
including processors (such
as X86, POWER, and ARM cores), system
communication links (such as PCIe, NVLinks, CAPI/OpenCAPI), 
special system accelerators (both
discrete ones such as FPGAs and
integrated ones such as Tensor Cores),
 libraries (such as CUDA and CuDNN), and
frameworks (such as TensorFlow, Caffee, and
PyTorch).

Instead of developing ad-hoc performance measurement 
solutions for each task, we decide to develop
a common system characterization and 
benchmarking infrastructure and tooling for all
of our tasks, providing some uniformity across 
the various tasks.

As time passes by, we find that this infrastructure 
is proven not only useful, but also
greatly boosted our productivity. A number
of interesting systems projects at \c3sr have
already benefited from having such an infrastructure
and tooling. Therefore, we decide to open source
this infrastructure so that other research teams 
interested in
systems work can also benefit from our work.
Hence the genesis of this \c3sr project: \myscope.

%% file: sections/2-design-philosophy.tex
\section{Design Philosophy}
\label{sec:intro}

This whitepaper describes the motivation and design of the initial release (v1.0) of the Scope benchmarking infrastructure.
Scope is designed around three primary goals:
\begin{itemize}
\item \textit{extensibility}: 
It should be easy for external groups to develop independent benchmarks without requiring centralized coordination with the Scope project.
This allows different teams to develop their own measurement tools to fit their own needs.
\item \textit{portability}:
The Scope infrastructure should support as many different systems as possible.
Though individual benchmarks may have specific requirements, Scope itself should not be a barrier to running benchmarks on a particular system.
Scope has been tested on POWER8/POWER9- and x86\_64-based systems, but should support any system that has a C++11 compiler and the CUDA~\cite{nvidia2018cuda} toolkit.
\item \textit{development silos}:
New groups of benchmarks should be able to be open- or closed-source.
Each group of benchmarks may have its own software dependencies, compiler feature requirements, or other necessities, and those requirements should not be globally propagated to all benchmark code.
This allows Scope to remain system-agnostic and useful to the widest possible audience.
\end{itemize}

\begin{figure}[htbp]
    \includegraphics[width=\linewidth]{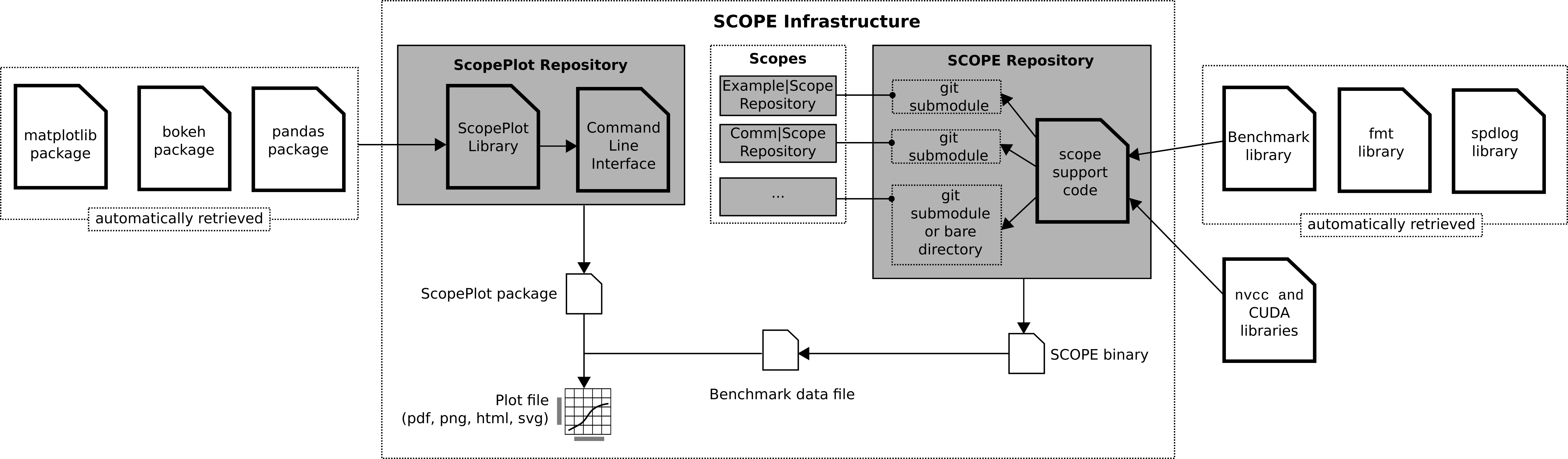}
    \caption{ 
Diagram of Scope architecture.
The scope infrastructure is divided into three components.
The Scope Repository
contains code to initialize and run registered scopes and contains the \texttt{main} function (the entry point) to the SCOPE binary.
Scopes are separate structured directories, usually git repositories, that contain the actual benchmark code. 
The ScopePlot repository holds Python code for implementing the ScopePlot python package.
Running the SCOPE binary produces a benchmark data file, which can be consumed by ScopePlot to produce plots of results.
    }
    \label{fig:scope-architecture}
\end{figure}

The Scope infrastructure consists of three kinds of software components.
First, the SCOPE repository (Section~\ref{sec:scope}) manages configuration and compilation and provides shared utility functions for the benchmarks.
Second, \textit{scopes} (Section~\ref{sec:scopes}) define groups of benchmarks, with their own optional dependencies and utilities.
Third, the ScopePlot project (Section~\ref{sec:scope-utilities}) provides a Python package for plotting and manipulating Scope results.
Figure~\ref{fig:scope-architecture} shows the relationship between Scope infrastructure components.

SCOPE and ScopePlot make heavy use of existing tools and libraries, notably Git~\cite{git2018} source control, CMake~\cite{kitware2018cmake} compilation configuration, the Google Benchmark library~\cite{google2018benchmark}, matplotlib~\cite{hunter2007matplotlib}, bokeh~\cite{bokeh2018}, and pandas\cite{mckinney2010data}.

\subsection{Licensing, Hosting, and Contributing}

The Scope infrastructure is free and open source, licensed under the Apache 2.0 license~\cite{apache22018}.
Scope welcomes contributions.
See \href{https://github.com/c3sr/scope/blob/master/CONTRIBUTING.md}{\texttt{CONTRIBUTING.md}} in the Scope source tree for up-to-date information about contributing.
Table~\ref{tab:host} lists the URLs for Scope infrastructure components.

\begin{table}[htbp]
\caption{Hosting Locations}
\begin{center}
\begin{tabular}{c c}
\hline
\textbf{Component} & \textbf{URL} \\ \hline
Scope     & \href{https://github.com/c3sr/scope}{\texttt{https://github.com/c3sr/scope}}       \\ \hline
ScopePlot & \href{https://github.com/c3sr/scope\_plot}{\texttt{https://github.com/c3sr/scope\_plot}} \\ \hline
\end{tabular}
\label{tab:host}
\end{center}
\end{table}

%% file: sections/3-scope.tex
\section{SCOPE Repository}
\label{sec:scope}

The SCOPE repository (\href{https://github.com/c3sr/scope}{\texttt{github.com/c3sr/scope}}) is the entry point for building and running benchmarks.
SCOPE is maintained as a Git repository to provide an open revision history and broad accessibility.
SCOPE itself does not contain any benchmark code; instead, it has the following responsibilities:
\begin{itemize}
\item retrieve benchmark code
\item configure the SCOPE binary compilation
\item fetch dependencies
\item provide common utilities
\item provide initialization hooks
\end{itemize}

\begin{figure}[htbp]
    \includegraphics[width=\linewidth]{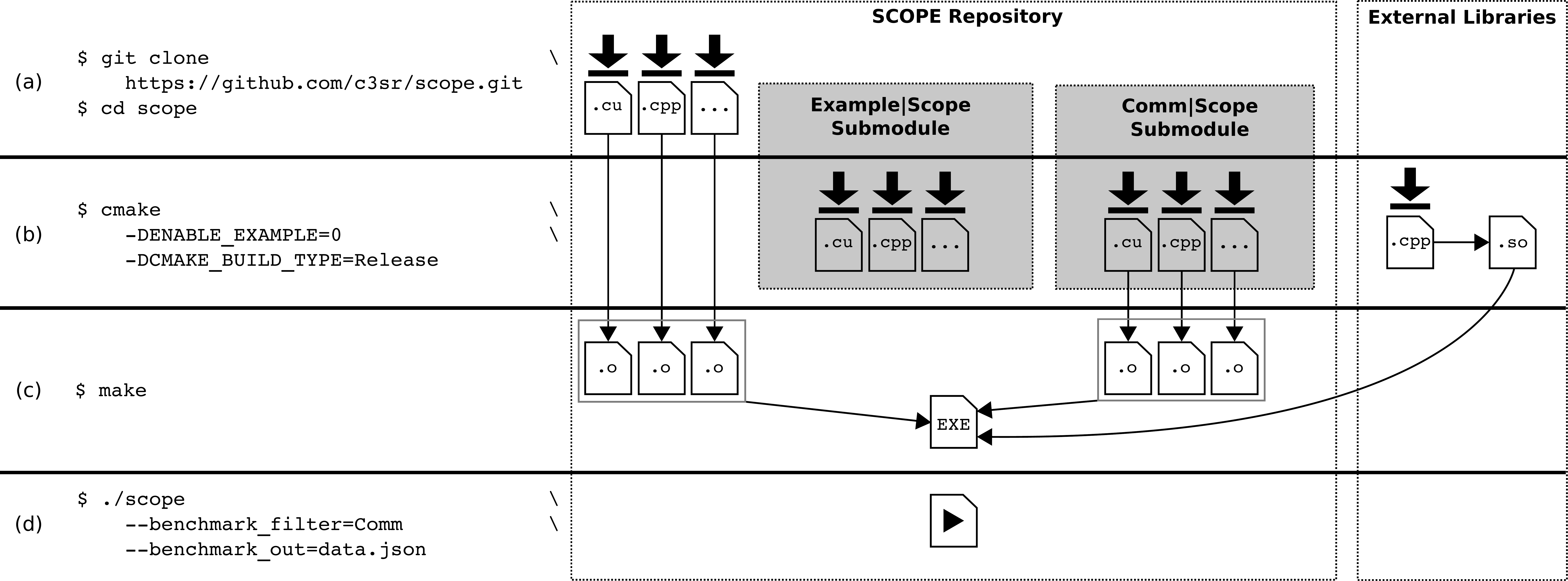}
    \caption{
Stages of building Scope.
(a), the download stage where SCOPE code is downloaded.
(b), the configuration stage where scopes are downloaded, scopes are enabled, and dependencies are downloaded and built.
(c) the build stage, where the SCOPE binary is produced.
(d) the run stage, where the benchmarks are run.
}
    \label{fig:scope-build}
\end{figure}



\subsection{Retrieving Code}
During the download stage, Figure~\ref{fig:scope-build}(a), the SCOPE source code is retrieved.

\subsection{Configuring the SCOPE Binary Compilation and Fetching Dependencies}
\label{sec:configure}

Since the SCOPE repository does not contain any benchmark code, a user will typically provide benchmark code by including additional scopes (Section~\ref{sec:scopes}) during the build process.
Scopes are semi-independent packages that contain benchmark code.
For development isolation, scopes are maintained as separate projects and are included in SCOPE as CMake-aware projects in the \texttt{scopes} directory of the SCOPE source tree.

SCOPE includes several completed scopes (Table~\ref{tab:scopes}) as Git submodules.
When scopes are added as submodules, it is easy to assicated each SCOPE release with a specific revision of benchmark code, and the correct version of the scopes can be automatically downloaded when downloading the whole scope project.
Other scopes can be manually added to this directory.

Scope uses CMake~\cite{kitware2018cmake} to configure the Scope binary compilation during the configuration stage, shown in Figure~\ref{fig:scope-build}(b).
CMake simplifies the process of supporting a variety of compilers and systems as well as fetching dependencies.
Through CMake, the Git submodules in SCOPE are downloaded and the fixed revisions are checked out.
SCOPE then invokes CMake's \texttt{add\_subdirectory} command on all directories in the \texttt{scopes} directory to add those scopes to the SCOPE binary build.
Each scope exports a CMake Object Library~\cite{kitware2018addlibraryobjectform} - a CMake target that represents a set of object files containing the implementation of that scope, as well as associated information about required dependencies for those object files.

Since benchmarks may only be compatible with particular systems, Scope allows conditional compilation of benchmarks.
By selectively including or excluding object libraries during configuration, Scope may selectively include benchmark code and associated dependencies.
Each scope's \texttt{CMakeLists.txt} may define options for disabling the scope.
For example,
\texttt{cmake -DENABLE\_EXAMPLE=ON \textit{scope-source-directory}}
directs CMake to include object files and dependencies from \examplescope (Section ~\ref{sec:example-scope}) when building the Scope binary.

Scope relies on Hunter~\cite{baratov2018hunter} for CMake to fetch dependencies.
Some dependencies, such as \texttt{spdlog}, are present in Scope so all benchmarks may used them.
Table~\ref{tab:scope-library-dependencies} shows the libraries included in Scope v1.0.0.
Other dependencies are only required for certain benchmarks.
For example, \commscope requires libnuma~\cite{libnuma2018} for pinning allocations or processes to memory regions on Linux systems, even though Scope itself does not.

Hunter downloads, configures, and builds the C++ library dependencies that SCOPE relies on.
Individual scopes may also use Hunter to retrieve their dependencies, or they may require the user to provide them in some other way (for example, as libraries present on the system).
If a scope uses Hunter, the scope's dependencies will only be retrieved if that scope was enabled in the configuration step.
All downloads through Hunter occur at the end of the configuration step, so that the required library and header files are present for the build.

\begin{table}[htbp]
    \caption{Scope Library Dependencies}
    \begin{center}
    \begin{tabular}{c c}
    \hline
    \textbf{Library} \\ \hline
    Google Benchmark v1.4.0~\cite{google2018benchmark} \\ \hline
    spdlog v0.16.3~\cite{spdlog2018} \\ \hline
    fmt v4.1.0~\cite{fmt2018} \\ \hline
    clara v1.1.4~\cite{horenovsky2018clara} \\ \hline
    \end{tabular}
    \label{tab:scope-library-dependencies}
    \end{center}
\end{table}

\subsection{Compiling SCOPE}

During the build stage (c), object files for all enabled scopes are produced and linked into a single binary.

In this step, every source file that was included in a scope submodule object library is compiled into an object file.
The source files that implement the Scope infrastructure helpers are also compiled into object files, and then all of the objects are linked together in a single step to form the scope binary.

\subsection{Running SCOPE}

Finally, running the produced SCOPE binary allows the user to select any subset of the included benchmarks and produce a data file at any location.

\subsection{Providing Common Utilities}
\label{sec:scope-common-utilities}

Scope provides a set of utilities that scopes may use.
The interfaces for these utilities are available to the scopes through C++ header files.
When scopes include those headers, the implementations will be available when all objects are linked together.

\begin{itemize}
\item The entire Google Benchmark library is provided to configure and register the benchmark code.
\item CUDA error checking is provided, as most extant benchmarks are CUDA benchmarks.
\item Logging is provided so that scopes may have a consistent output mechanism.
\item C++ functions for declaring new command line options.
\item C++ function for executing pre-benchmark initialization code.
\item Convenience CMake functions for integrating with SCOPE.
\end{itemize}

Scope also provides a \texttt{tools/generate\_sugar\_files.py} python script for generating Sugar-compatible\cite{baratov2018sugar} CMake files in each scope source tree.
Scope uses Sugar to read the \texttt{sugar.cmake} files, which tell CMake where the CUDA and C++ source files for Scope and the benchmarks are.
This script is provided so that the \texttt{sugar.cmake} files can be quickly regenerated during development of each scope.
The script will create \texttt{sugar.cmake} files that export the CMake variables described in Table~\ref{tab:sugar-variables}.
\texttt{<VAR>} is replaced by the string passed to the ``--var'' flag.

\begin{table}[htbp]
    \caption{CMake variables created by \texttt{sugar.cmake} files}
    \begin{center}
    \begin{tabular}{c c}
    \hline
    \textbf{Library} & \textbf{Source} \\ \hline
    \texttt{<VAR>\_SOURCES} & C/C++ source files\\ \hline
    \texttt{<VAR>\_HEADERS} & C/C++ header files\\ \hline
    \texttt{<VAR>\_CUDA\_SOURCES} & CUDA source files \\ \hline
    \texttt{<VAR>\_CUDA\_HEADERS} & CUDA header files \\ \hline
    \end{tabular}
    \label{tab:sugar-variables}
    \end{center}
\end{table}

\subsection{Providing CMake Functions}

SCOPE provides three functions to help scopes integrate with the SCOPE CMake build. 
\texttt{Scope\_add\_library} is a wrapper around \texttt{add\_library}, which also sets the \texttt{SCOPE\_NEW\_TARGET} variable so that SCOPE can link against the CMake Object Library defined in the scope.
\texttt{Target\_include\_scope\_directories} causes SCOPE's utility include directories to be added to the compilation of the scope so that the SCOPE utilities can be used.
SCOPE also provides \texttt{scope\_status}, \texttt{scope\_warning}, and \texttt{scope\_fatal}, which scopes may use in their \texttt{CMakeLists.txt} files to print messages that will be visible during configure time.

\subsection{Initialization Hooks and Command-line Options}

SCOPE provides the ability for benchmarks to run arbitrary initialization when the binary is executed.
Scopes may register \texttt{clara::Opt}s to create new command line arguments accepted by the SCOPE binary.
Scopes may also register arbitrary code to be executed before command line arguments are parsed, or after arguments are parsed, but before any benchmarks are executed.
Though these routines, benchmarks may do any unguided or user-directed initialization desired before benchmark execution.

%% file: sections/4-scope-submodules.tex
\section{Design of SCOPE Submodules}
\label{sec:scopes}

SCOPE does not include any benchmark code --- all benchmarks are provided through individual \textit{scopes}.
Scopes are structured directories in the Scope source tree, included in Scope through the CMake \texttt{include\_directory} command.
As of this writing, eight scopes are either completed or in development, to measure various aspects of system compute and communication performance at different levels of abstraction.
Table~\ref{tab:scopes} describes the different scopes that are under development.

\begin{table}[htbp]
    \caption{Completed or in-progress scopes}
    \begin{center}
    \begin{tabular}{c c c l}
    \hline
    \textbf{Abstraction} & \textbf{Name} & \textbf{Status} & \textbf{Description} \\ \hline
    Hardware & TCU$\vert$Scope & Completed & Nvidia GPU tensor cores \\
    \hline 
    \multirow{3}{*}{Data Transfer} & Comm$\vert$Scope & Released & Nvidia CPU-GPU communication~\cite{pearson2018commscope} \\
    & I/O$\vert$Scope & In Progress & Disk I/O operations \\
    & NCCL$\vert$Scope & In Progress & Nvidia's NCCL library \\
    \hline
    \multirow{4}{*}{Compute} & cdDNN$\vert$Scope & Released & Neural-network operations \\
    & Instr$\vert$Scope & In Progress & Instruction latencies and throughput \\
    & Histo$\vert$Scope & In Progress & Nvidia GPU histogramming \\
    & LinAlg$\vert$Scope & In Progress & Linear algebra operations \\
    \hline
    \end{tabular}
    \label{tab:scopes}
    \end{center}
\end{table}


\subsection{Defining a CMake Object Library}

Each scope needs to include a \texttt{CMakeLists.txt} at the top level of its directory.
This allows it to be included in the Scope configuration through the \texttt{include\_directory} command in the Scope \texttt{CMakeLists.txt}.
Fundamentally, each scope only needs to define a CMake object library target though the object form of the \texttt{add\_library}  command~\cite{kitware2018addlibraryobjectform}.
\examplescope defines the \texttt{example\_scope} target.
Scope suggests using Sugar and the provided tooling (Section~\ref{sec:scope-common-utilities}) to automate the process of providing source files to the \texttt{scope\_add\_library} command.
The scope target may have its own dependencies or other constraints.
\examplescope marks itself as requiring C++11 through the \texttt{target\_compile\_features} CMake command.
This tells CMake that objects in \examplescope should be built with C++11 support in the compiler.
\examplescope also adds include directories, a CUDA language standard, and a requirement to link against the Google Benchmark library to the \texttt{example\_scope} target.
These requirements will be propagated to the entire Scope build as required.

\subsection{Integration with Scope through Git Submodules}
For development isolation, scopes are maintained as independent directories.
If the scope is maintained as a Git repository, the scope can have its own versioning and revision history.
Additionally, this allows the scope to be included in the SCOPE repository as a Git submodule.
Git submodules can be automatically downloaded alongside Scope when Scope is downloaded, and pinned to the appropriate version.

\subsection{Example Scope}
\label{sec:example-scope}

The Scope project provides a template scope \examplescope~\cite{pearson2018examplescope} that demonstrates how a new scope can be structured.
\examplescope is available at \texttt{\url{https://github.com/c3sr/example_scope}}.
\examplescope demonstrates the following required or suggested structures:
\subsubsection{\texttt{CMakeLists.txt} (required) and Sugar (optional)}
This file defines a CMake object library and uses Sugar to parse the \texttt{sugar.cmake} files in the \examplescope source tree.

\subsubsection{Code Structure (optional)}
\examplescope places all of its source files in the \texttt{src} directory.

\subsubsection{Benchmark Library (required)}
All benchmarks are registered through the Benchmark library.
This enables the Scope binary to filter, run, and report results in a consistent way.

\subsubsection{Documentation (optional)}
\examplescope contains Markdown files describes each benchmark, the algorithm, and implementation in \texttt{docs} in its source tree.

\subsubsection{Initialization and Command Line Flags (optional) }
\examplescope uses \texttt{clara::Opt}s to declare two new command-line arguments, and uses the SCOPE initialization hooks to cause SCOPE to exit during initialization if those options are used.
Initialization code is placed in \texttt{src/init}.

%% file: sections/5-scope-utilities.tex
\section{Scope Utilities}
\label{sec:scope-utilities}

\subsection{ScopePlot Python Package}

ScopePlot is a python package available to help plot and manipulate results in the JSON files produced by SCOPE.
The Google Benchmark-formatted JSON files (hereafter referred to as JSON files) produced by Scope are unmodified from the format produced by the Google Benchmark library, so ScopePlot is compatible with other tools that use that library.
ScopePlot is freely available on the Python Package Index (PyPI) at \texttt{\url{https://pypi.org/project/scope-plot/}}.
ScopePlot is also open-source, hosted at \texttt{\url{https://github.com/rai-project/scope_plot}}.
ScopePlot uses Python's distutils to manage installation.
When ScopePlot is installed, it provides the \texttt{scope\_plot} binary.
The rest of this section describes notable \texttt{scope\_plot} subcommands.

\subsubsection{\texttt{spec} Subcommand}

\texttt{scope\_plot spec} generates an arbitrary plot from a YAML~\cite{yaml2018} specification file (hereafter referred to as a \textit{spec} file).
Examples of valid specification files can be found in the ScopePlot source tree.
The specification file controls the plot type (line with error bars, bar plot, linear regression plot with error bars), the source JSON file for each data series, filters to extract the desired data from the JSON file, per-series data transformations, and plot styling and formatting.

\begin{figure}[htbp]
\centering
    \includegraphics[width=0.5\linewidth]{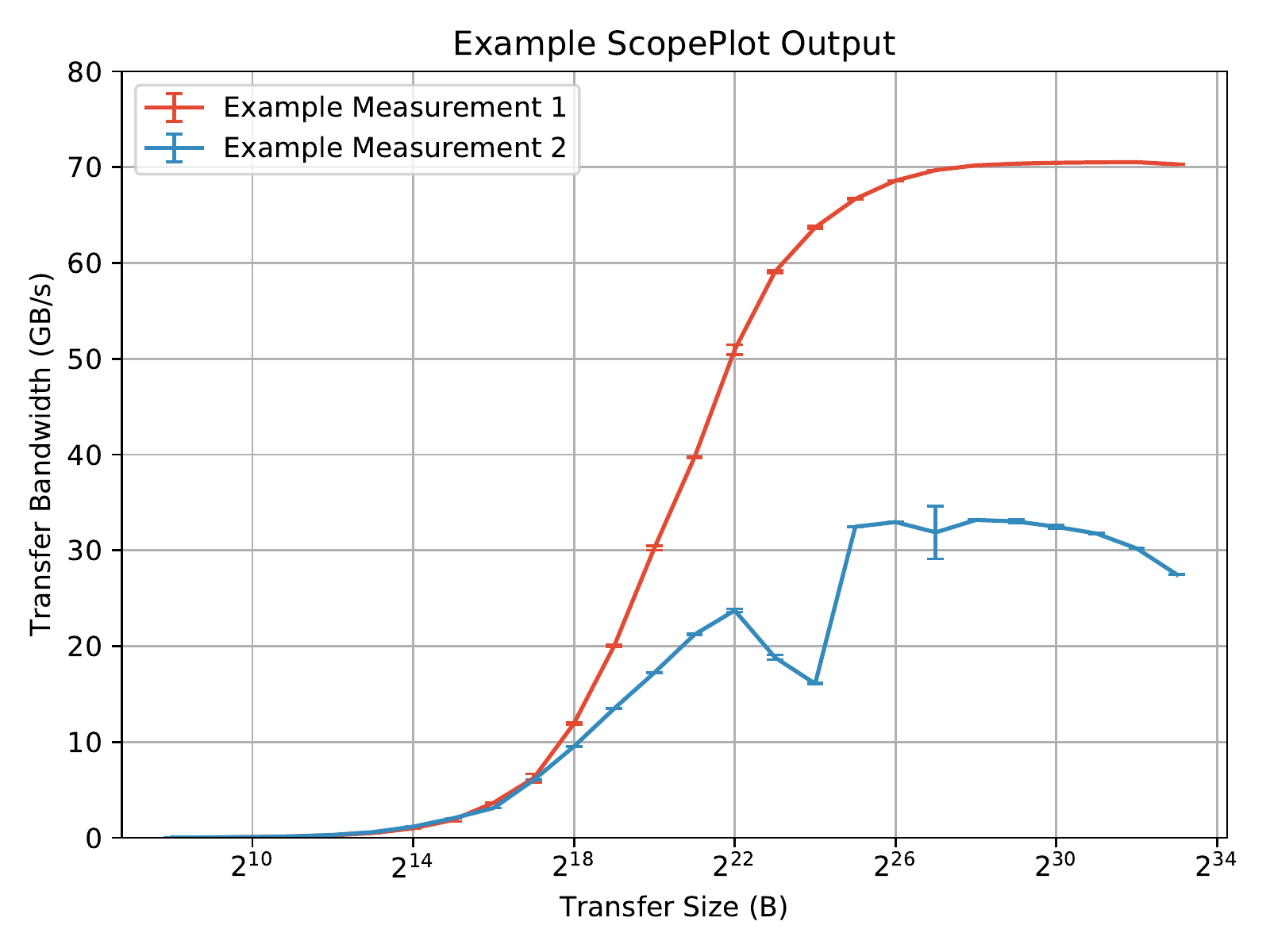}
    \caption{ 
Example line plot with error bars generated from ScopePlot.
    }
    \label{fig:plot-example}
\end{figure}

\subsubsection{\texttt{deps} Subcommand}

\texttt{scope\_plot deps} can be used to help integrate the \texttt{scope\_plot} command with GNU Makefiles.
Similar to how make can be used to partially recompile code when certain files have changed, make can be used to regenerate plots when the underlying Benchmark JSON files have been updated.
\texttt{deps} scans a \textit{spec} file and emits the paths of the benchmark JSON files it depends on in make format, so that make target dependencies can be automatically generated as part of a build process.

\subsubsection{\texttt{bar} Subcommand}
\texttt{scope\_plot bar} generates a bar plot from a Benchmark JSON file.
It has a subset of the functionality of the \texttt{spec} subcommand, without requiring a full \textit{spec} file.
Command line options allow the user to specify the fields used for the x- and y-axis data, and the plot title.

\subsubsection{\texttt{cat} Subcommand}
\texttt{scope\_plot cat} is inspired by the Linux/Unix ``cat'' command, but specialized to Google Benchmark JSON files.
When passed one or more Benchmark JSON files, it concatenates the \texttt{benchmarks} field of those files and dumps the content to the standard output stream.
In this way, it preserves the structure of the JSON files when they are concatenated, where the standard ``cat'' would simply append the JSON contents together, yielding a malformed result.

\subsubsection{\texttt{filter\_name} Subcommand}
\texttt{scope\_plot filter\_name} filters the benchmark outputs in the Benchmark JSON file to only keep benchmarks with a name that matches a provided regular expression.

\subsubsection{Using ScopePlot as a Library}
ScopePlot may also be used as a python library to develop other JSON file manipulation and analysis tools.
ScopePlot has an object model for JSON files and various methods for filtering them and converting them to pandas DataFrames.

\subsection{Scope Docker Images}

\begin{table*}[t]
    \caption{Dockerfiles and Docker Images}
    \begin{center}
    \begin{tabular}{c c c}
    \hline
    \textbf{Dockerfile} & \textbf{Docker Hub Image} & Description \\ \hline
    \texttt{amd64.cuda75.Dockerfile}   & c3sr/scope:amd64-cuda75-\textit{tag} & x86\_64, CUDA 7.5, \\ \hline
    \texttt{amd64.cuda80.Dockerfile}   & c3sr/scope:amd64-cuda80-\textit{tag} & x86\_64, CUDA 8.0, \\ \hline
    \texttt{amd64.cuda92.Dockerfile}   & c3sr/scope:amd64-cuda92-\textit{tag} & x86\_64, CUDA 9.2, \\ \hline
    \texttt{ppc64le.cuda80.Dockerfile} & c3sr/scope:amd64-cuda80-\textit{tag} & POWER, CUDA 8.0, \\ \hline
    \texttt{ppc64le.cuda92.Dockerfile} & c3sr/scope:amd64-cuda92-\textit{tag} & POWER, CUDA 9.2, \\ \hline
    \end{tabular}
    \label{tab:dockerfiles}
    \end{center}
\end{table*}

Scope includes several Docker~\cite{merkel2014docker} images, which can be used to run benchmarks.
This allows users on supported platforms to run pre-packaged versions of the benchmark code without having to compile or configure it.
Table~\ref{tab:dockerfiles} lists the Docker files and corresponding Docker images publicly available on the Docker Hub.

%% file: sections/6-dev-maint.tex
\section{SCOPE Development and Maintenance}

\begin{figure}[tbp]
    \includegraphics[width=\linewidth]{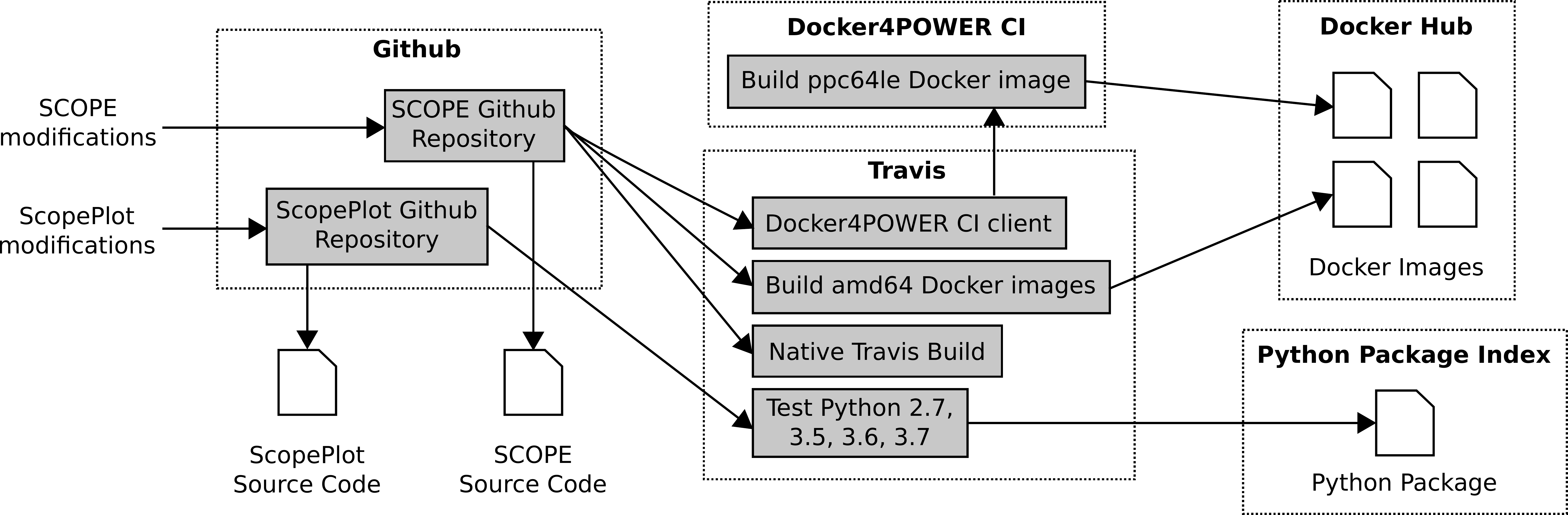}
    \caption{
SCOPE and ScopePlot development flow.
Once source code is pushed to Github, it is immediately available there.
Travis-CI is used to build Scope, and also start Docker4POWER CI jobs to build POWER Docker images.
Travis-CI is also used to test ScopePlot.
After successful builds or tests, the relevant artifacts are deployed to Docker Hub and the Python Package Index.
}
    \label{fig:ci}
\end{figure}

Scope and ScopePlot rely on continuous integration for testing and building Docker images.
The continuous integration system is centered around Travis-CI~\cite{travis2018}.
Whenever modifications to Scope or ScopePlot are pushed to GitHub, Travis-CI starts a series of parallel jobs.
Figure~\ref{fig:ci} summarizes the continuous integration flow for Scope and ScopePlot.

\subsection{Scope}

Whenever a push is made to the Scope repository, Travis-CI is configured to start a series of parallel builds:
\begin{itemize}
    \item An x86-64 CUDA 9.2 build.
    \item An x86-64 Docker CUDA 7.5 build.
    \item An x86-64 Docker CUDA 8.0 build.
    \item An x86-64 Docker CUDA 9.0 build.
\end{itemize}
Each of these builds incorporate \commscope and \examplescope, the two completed scopes at the time of writing.

All of Travis' build hardware is x86-64, so Travis is directly used to create x86-64-compatible docker images.
To generate POWER-compatible Docker images, Scope uses rai~\cite{dakkak2017rai}, a separate job submission system.
Two additional Travis jobs are started, each of which simply submit the POWER builds to rai on Oregon State University's PowerCI~\cite{osu2018powerci} infrastructure.
\begin{itemize}
    \item POWER Docker CUDA 8.0 build.
    \item POWER Docker CUDA 9.2 build.
\end{itemize}

If these Docker builds individually succeed, they are pushed to Docker hub to be immediately available to the public.
An image corresponding to each tag is retained indefinitely.
Furthermore, the most recent commit on each branch is available.

\subsection{ScopePlot}

Whenever a push is made to the ScopePlot repository, Travis-CI is configured to test the ScopePlot package against Python 2.7, 3.4, 3.5, 3.6, and 3.7.
If all of those tests pass, and the commit has a corresponding tag, a new version of ScopePlot is made available on the Python Package Index (PyPI) for installation with \texttt{pip install scope\_plot}.

%% file: sections/7-conclusion.tex
\section{Conclusion}
\label{sec:conclusion}

The Scope project arose out of a desire to lower the barrier to entry for system benchmarking in the IBM / University of Illinois Urbana-Champaign Center for Cognitive Computing Systems Research (\c3sr).
Scope does this by incorporating common libraries and providing convenience functions for writing new system benchmarks, easily supporting compilation on x86 and POWER platforms, and by providing a command-line tool for managing and plotting results.
Scope is free and open-source, and welcomes contributors and collaborators.